\documentclass[]{elsart}

\usepackage{natbib}

\usepackage[english]{babel}
\usepackage{longtable}

\usepackage{graphics}
\usepackage{graphicx}
\usepackage{epsfig}

\usepackage{amssymb}

\begin{document}

\begin{frontmatter}


\title{CCD and photon-counting photometric observations of asteroids
carried out at Padova and Catania observatories}
\author{D. Gandolfi$^{1,3}$,}
\author{M. Cigna$^{2}$,}
\author{D. Fulvio$^{1,2}$\corauthref{corr. author}}
\corauth[corr. author]{Corresponding author. Tel:+39 095 7332325; Fax:+39 095 330592; email address: dfu@oact.inaf.it}
\author{and C. Blanco$^{1}$}

\address{$^{1}$Department of Physics and Astronomy, Catania University, via S.Sofia 64, I-95123, Catania, Italy}
\address{$^{2}$INAF-Catania Astrophysical Observatory, via S. Sofia 78, I-95123, Catania, Italy}
\address{$^{3}$Th\"uringer Landessternwarte Tautenburg, Sternwarte 5, D-07778 Tautenburg, Germany}

\begin{abstract}

We present the results of observational campaigns of asteroids performed
at Asiago Station of Padova Astronomical Observatory and at M.G. Fracastoro
Station of Catania Astrophysical Observatory, as part of the large research
programme on Solar System minor bodies undertaken since 1979 at the Physics
and Astronomy Department of Catania University. Photometric observations of
six Main-Belt asteroids (27~{\it Euterpe}, 173~{\it Ino}, 182~{\it Elsa},
539~{\it Pamina}, 849~{\it Ara}, and 984~{\it Gretia}), one Hungaria
(1727~{\it Mette}), and two Near-Earth Objects (3199~{\it Nefertiti} and
{\it 2004~UE}) are reported. The first determination of the synodic rotational
period of {\it 2004~UE} was obtained. For 182~{\it Elsa} and 1727 {\it~Mette}
the derived synodic period of $80.23\pm0.08~h$ and $2.981\pm0.001~h$, respectively,
represents a significant improvement on the previously published values. For
182~{\it Elsa} the first determination of the $H-G$ magnitude relation is also
presented.

\end{abstract}


\end{frontmatter}

\section{Introduction}
\label{Sec:Introduction}

Photometric observations of asteroids, through the lightcurves collection,
allow us to determine their rotational parameters (synodic rotational
period, spin axis orientation, sense of rotation, etc. ) and physical
properties (shape, colour indexes, surface dishomogeneity, taxonomic
class, etc.). The knowledge of the rotational characteristics is important to
understand the collisional evolution of single minor planets, of families
and of the whole asteroidal population. It is also essential to check the
action of the so-called Yarkovsky-O'Keefe-Radzievski-Poddak (YORP) effect on smaller
asteroids. It seems to become clear that the spin characteristics of
super-fast and slow rotators are the result of the torque from solar radiation
and re-radiation \citep{Harris06} with the collisional processes determining
the excitation of the tumbling motion on slow rotators.

The results reported in this paper are part of the large photometric observational
programme of asteroids undertaken since 1979 at the Physics and Astronomy Department
of Catania University \citep{Blanco79, DiMartino84, DiMartino94, Blanco98, Riccioli01}. 
The aim of this long-term programme is to increase the number of asteroids with 
well-known rotational parameters and consequently improve the database necessary for 
the investigation of the minor planet evolutionary history. Special attention has been 
devoted to the asteroids with observational constraints and to those with few known 
lightcurves, in order to obtain the minimum number needed to apply pole coordinates 
and shape computational methods \citep{Blanco98}. Since the study of Near-Earth 
Objects (NEOs) is important to improve the knowledge of inner Solar System mechanics and 
physics (resonance mechanisms, smaller bodies dynamics, connection with meteorites, etc.), 
we have recently expanded our programme to include these objects.

In this paper we present new photometric observations of six Main-Belt asteroids,
27~{\it Euterpe}, 173~{\it Ino}, 182~{\it Elsa}, 539~{\it Pamina}, 849~{\it Ara}, and
984~{\it Gretia}, one Hungaria object, 1727~{\it Mette}, and two NEOs, 3199~{\it Nefertiti}
and {\it 2004~UE}.

The values of the synodic rotational period, lightcurve amplitudes, and $B-V$ colour index
of almost all the observed asteroids are presented. For 182~{\it Elsa} and 1727~{\it Mette}
the rotational periods we derived represent a significant improvement on the previously
published values. For 182~{\it Elsa} also the {\it H-G} magnitude relation \citep{Bowell89} was
obtained. The first observed lightcurve of the small NEO {\it 2004~UE} might
indicate an elongated shape and possible smooth surface.

The outline of the paper is as follows: we describe our observations and data reduction in
Section~\ref{Sec:PhotObs-DataRed}; the results and comments on each single object are reported in 
Section~\ref{Sec:Results}.

\section{Photometric observations and data reduction}
\label{Sec:PhotObs-DataRed}

The photometric observations reported in this paper were carried out
at two different observatories using different telescopes and
instruments.

CCD imaging photometry was obtained during three observing runs at
the Asiago Station of Padova Astronomical Observatory (hereafter PD station),
by using the 67/92-cm Schmidt telescope. Unfiltered CCD photometry was
performed with the SCAM-1 camera in January and February 2003 with mostly
clear and stable weather conditions (mean seeing $\approx 2.0\,^{\prime\prime}$).
The SCAM-1 camera hosts a front illuminated $2048\times2048$ LORAL CCD
with a $15\,\mu m$ pixel-size. Taking into account the telescope plate scale
($95.9\,^{\prime\prime}$/mm), the resulting projected sky area was about
$49\,^{\prime}\times49\,^{\prime}$ with an angular resolution of about
$1.44\,^{\prime\prime}$/pixel. Further CCD photometric observations were
carried out using the same telescope equipped with the ITANET camera
\citep{Blanco04, Gandolfi06}, hosting a front illuminated
$2048\times2048$ Kodak-KAF 4200 CCD, with $9\,\mu m$ pixel-size. The
observations were performed through $V$ and $R$ Johnson filters in April 2003
and November 2004, under photometric weather conditions with seeing varying
between $1.5$ and $2.0\,^{\prime\prime}$. Taking into account
the telescope plate scale and the detector pixel-size, the resulting field of view
turned out to be $29\,^{\prime}\times\,29\,^{\prime}$ with an angular
resolution of $0.89\,^{\prime\prime}$/pixel. The exposure time during both
campaigns varied from 2 to 5 minutes, depending on the asteroid magnitude.

CCD image pre-reduction, including bias, dark, and flat-field correction,
was made with standard IRAF\footnote{IRAF is distributed
by the National Optical Astronomy Observatory, which is operated by
the Association of Universities for Research in Astronomy (AURA), inc.,
under cooperative agreement with the National Science Foundation.} routines.
Nightly twilight flat-fields were used to correct images for optical
vignetting, dust shadow, and pixel-to-pixel sensitivity variation. In order
to maximize the asteroid signal-to-noise ratio (especially in cases of elongated
images of fast moving asteroids), elliptical aperture photometry was
performed using the SExtractor software \citep{Bertin96}. A set
of comparison stars, each having magnitudes similar to that of the observed
asteroid, was selected along the object nightly path. From this set of
stars, the non-variable star with the highest signal-to noise ratio was chosen
as the nightly comparison star. Due to a filter wheel technical problem no
fields of standard star were observed during both runs. Therefore only
differential photometry was obtained from CCD imaging data.

Photoelectric photometry was carried out in November 1993, December 1996,
and July, September, October and November 2004 at the 91-cm Cassegrain telescope
of M.G. Fracastoro Station of INAF-Catania Astrophysical Observatory
(hereafter CT station). The observations were performed with the Johnson
$B$ and $V$ standard filters. A 1.5-mm diameter diaphragm, limiting the
telescope field to about $22\,^{\prime\prime}$, and a cooled photon-counting
single-head photometer equipped with an EMI\,9893QA/350 photomultiplier were used.
Nearby solar spectral type comparison stars were selected along
the asteroid path to neglect the second-order chromatic effects of
atmospheric extinction. The $B$ and $V$ linear extinction coefficients
were derived each night through the comparison stars. In order to determine the
transformation coefficients to the Johnson standard system, several standard
stars, selected from \citet{Mermilliod97}, \citet{Blanco68}, and \citet{Landolt92}, 
were also observed during each night. The observing strategy and data reduction 
were the same as those already adopted during previous observational campaigns of 
asteroids \citep{DiMartino94}.

Both for CCD and photoelectric data, the final error of the single
measurements is on average between $\sim0.01$ and $0.02$~mag. On the
basis of the asteroid aspect data, the time corresponding to each
data point was corrected for light-travel time and, when determined,
the standard $V$ magnitudes were also reduced to the unit geocentric
and heliocentric distances (\,$V(1,\alpha)$\,). The value of the synodic
rotational period, the composite lightcurve, the mean reduced magnitude
$\bar{V}(1,\alpha)$, and the nightly magnitude shifts were obtained by
applying the Fourier analysis, as described in \citet{Harris89}.

The nightly aspect data for each asteroid observed at PD and CT station
are listed in Table~\ref{Table1}. The first column reports the mean Universal Time (UT)
of the nightly observational interval, rounded to a hundredth of a day.
The other columns give the heliocentric ecliptic longitude
and latitude, asteroid-Sun ($r$) and asteroid-Earth ($\Delta$) distances,
and solar phase angle ($\alpha$) at the mean-time of each observing night.
For the asteroids observed at CT station, the value of the mean reduced
magnitude $\bar{V}(1,\alpha)$ are also reported. The filters used and
observatory's site are listed in the last two columns.

\clearpage
\begin{scriptsize}
\begin{longtable}{lcccccccc}

\caption{Aspect data of the asteroids. The heliocentric ecliptic longitude ($\lambda$) and latitude ($\beta$),
         asteroid-sun ($r$) and asteroid-Earth ($\Delta$) distances, and solar phase angle ($\alpha$) are referred to the
         nightly mean UT listed in the first column. When derived, the mean reduced magnitude (\,$\bar{V}(1,\alpha)$\,)
         is also reported. The filters used and observatory's site are listed in the last two columns.}\\

\hline
\hline
\noalign{\smallskip}
\hspace{0.3 cm}Mean UT  &$\lambda$\,(2000)&$\beta$\,(2000)&   r   & $\Delta$ &  Phase Angle  & $\bar{V}(1,\alpha)$ &  Filters & Obs. \\

(yy/mm/dd)              &     (degree)    &   (degree)    &  (AU) &   (AU)   &   (degree)    &        (mag)        &   Used   &      \\
\noalign{\smallskip}
\hline
\endfirsthead

\caption{Continued.}\\
\hline
\hline
\noalign{\smallskip}
\hspace{0.3 cm}Mean UT  &$\lambda$\,(2000)&$\beta$\,(2000)&   r   & $\Delta$ &  Phase Angle  & $\bar{V}(1,\alpha)$ &  Filters & Obs. \\

(yy/mm/dd)              &     (degree)    &   (degree)    &  (AU) &   (AU)   &   (degree)    &        (mag)        &   Used   &      \\
\noalign{\smallskip}
\hline
\noalign{\smallskip}
\noalign{\smallskip}
\endhead

\hline
\endfoot

\hline
\endlastfoot

{\bf 27 Euterpe}        &               &            &       &          &               &                     &          &      \\

1993 Nov 13.89          &     21.21     &    -1.52   & 2.150 &  1.390   &    21.119     &         --          &   BV     &  CT  \\

1993 Nov 14.88          &     21.53     &    -1.52   & 2.148 &  1.398   &    21.435     &         --          &   BV     &  CT  \\

2004 Nov 06.04          &     44.63     &    -1.22   & 2.038 &  1.048   &     1.349     &        7.062        &   BV     &  CT  \\

2004 Nov 07.00          &     44.97     &    -1.21   & 2.037 &  1.047   &     1.152     &        7.087        &   BV     &  CT  \\

2004 Nov 09.06          &     45.72     &    -1.20   & 2.034 &  1.045   &     1.610     &        7.079        &   BV     &  CT  \\

{\bf 173 Ino}           &               &            &       &          &               &                     &          &      \\

1996 Dec 07.05          &    101.99     &   -10.39   & 2.575 &  1.773   &    15.398     &        8.820        &   BV     &  CT  \\

1996 Dec 08.02          &    102.22     &   -10.35   & 2.577 &  1.767   &    15.090     &        8.811        &   BV     &  CT  \\

1996 Dec 16.07          &    104.15     &   -10.02   & 2.595 &  1.722   &    12.376     &        8.690        &   BV     &  CT  \\

2004 Jul 18.00          &    299.41     &     6.99   & 2.513 &  1.513   &     5.319     &        8.435        &   BV     &  CT  \\

2004 Jul 19.03          &    299.66     &     6.93   & 2.511 &  1.509   &     5.085     &        8.416        &   BV     &  CT  \\

2004 Jul 20.04          &    299.92     &     6.88   & 2.509 &  1.506   &     4.888     &        8.423        &   BV     &  CT  \\

2004 Jul 20.98          &    300.15     &     6.83   & 2.506 &  1.503   &     4.740     &        8.398        &   BV     &  CT  \\

2004 Jul 22.04          &    300.41     &     6.77   & 2.504 &  1.500   &     4.616     &        8.417        &   BV     &  CT  \\

{\bf 182 Elsa}          &               &            &       &          &               &                     &          &      \\

2004 Sep 09.02          &      2.64     &    -1.94   & 2.106 &  1.172   &    13.829     &        9.875        &   BV     &  CT  \\

2004 Sep 10.02          &      2.98     &    -1.94   & 2.104 &  1.165   &    13.372     &        9.927        &   BV     &  CT  \\

2004 Sep 11.04          &      3.33     &    -1.94   & 2.102 &  1.158   &    12.897     &        9.829        &   BV     &  CT  \\

2004 Sep 12.04          &      3.67     &    -1.95   & 2.100 &  1.152   &    12.425     &        9.772        &   BV     &  CT  \\

2004 Sep 13.08          &      4.02     &    -1.95   & 2.099 &  1.145   &    11.924     &        9.847        &   BV     &  CT  \\

2004 Sep 13.98          &      4.33     &    -1.95   & 2.097 &  1.140   &    11.489     &        9.589        &   BV     &  CT  \\

2004 Sep 15.06          &      4.70     &    -1.95   & 2.095 &  1.134   &    10.953     &        9.835        &   BV     &  CT  \\

2004 Sep 16.06          &      5.05     &    -1.96   & 2.094 &  1.128   &    10.453     &        9.818        &   BV     &  CT  \\

2004 Sep 21.02          &      6.76     &    -1.97   & 2.086 &  1.104   &     7.889     &        9.699        &   BV     &  CT  \\

2004 Oct 12.83          &     14.43     &    -2.00   & 2.053 &  1.064   &     5.331     &        9.622        &   BV     &  CT  \\

2004 Oct 17.94          &     16.26     &    -2.00   & 2.046 &  1.072   &     8.163     &        9.717        &   BV     &  CT  \\

{\bf 539 Pamina}        &               &            &       &          &               &                     &          &      \\

2004 Sep 11.00          &      1.24     &     6.80   & 2.162 &  1.213   &    11.907     &         --          &   BV     &  CT  \\

2004 Sep 12.03          &      1.59     &     6.80   & 2.161 &  1.208   &    11.477     &         --          &   BV     &  CT  \\

{\bf 849 Ara}           &               &            &       &          &               &                     &          &      \\

2004 Sep 15.90          &    339.39     &    18.29   & 2.672 &  1.789   &    12.597     &        8.941        &   BV     &  CT  \\

2004 Sep 20.85          &    340.63     &    18.15   & 2.679 &  1.819   &    13.488     &        8.945        &   BV     &  CT  \\

{\bf 984 Gretia}        &               &            &       &          &               &                     &          &      \\

2003 Feb 01.80          &     50.45     &    9.07    & 2.353 &  2.424   &    23.748     &         --          &    --    &  PD  \\

2003 Feb 06.80          &     51.92     &    9.04    & 2.360 &  2.488   &    23.275     &         --          &    --    &  PD  \\

2003 Feb 07.81          &     52.22     &    9.04    & 2.362 &  2.501   &    23.172     &         --          &    --    &  PD  \\

2003 Feb 09.79          &     52.80     &    9.03    & 2.364 &  2.527   &    22.963     &         --          &    --    &  PD  \\

{\bf 1727 Mette}        &               &            &       &          &               &                     &          &      \\

2003 Jan 31.11          &    136.42     &    1.40    & 1.723 &  0.750   &     7.676     &         --          &    --    &  PD  \\

2003 Feb 06.02          &    138.86     &    2.42    & 1.729 &  0.747   &     4.246     &         --          &    --    &  PD  \\

2003 Feb 06.96          &    139.25     &    2.59    & 1.730 &  0.747   &     3.980     &         --          &    --    &  PD  \\

2003 Apr 01.88          &    161.00     &   11.18    & 1.793 &  1.091   &    29.440     &         --          &     R    &  PD  \\

2003 Apr 04.93          &    162.22     &   11.61    & 1.797 &  1.122   &    30.120     &         --          &     V    &  PD  \\

2003 Apr 05.96          &    162.62     &   11.75    & 1.799 &  1.133   &    30.330     &         --          &     R    &  PD  \\

2003 Apr 07.93          &    163.40     &   12.03    & 1.801 &  1.153   &    30.706     &         --          &     R    &  PD  \\

2003 Apr 08.88          &    163.78     &   12.16    & 1.802 &  1.163   &    30.876     &         --          &     V    &  PD  \\

{\bf 3199 Nefertiti}    &               &            &       &          &               &                     &          &      \\

2003 Feb 01.11          &    150.47     &    6.18    & 1.647 &  0.792   &    24.738     &         --          &     --   &  PD  \\

2003 Feb 07.13          &    152.66     &    4.78    & 1.669 &  0.765   &    20.176     &         --          &     --   &  PD  \\

{\bf 2004UE}            &               &            &       &          &               &                     &          &      \\

2004 Nov 09.08          &     47.86     &   -0.08    & 1.010 &  0.025   &    39.128     &         --          &     R    &  PD  \\

\label{Table1}
\end{longtable}
\end{scriptsize}

\section{Results}
\label{Sec:Results}

With the exception of Figure~\ref{Fig:Elsa-HG}, where the $\bar{V}(1,\alpha)$ vs.
phase angle of 182 {\it Elsa} is plotted, the composite lightcurves are presented in
Figures~\ref{Fig:Euterpe-LC}-\ref{Fig:2004UE-LC}. Different symbols refer to different
observing nights. Only for the objects observed at CT station, the magnitude shifts,
applied to the single night lightcurve to obtain the composite one \citep{Harris89},
are reported in brackets near the observing date. The rotational phases were computed
according to the synodic rotational period reported in each Figures and Table~\ref{Table2}. The
period reliability codes \citep{Harris83}, lightcurve amplitude, and mean $B-V$
colour index, as well as the Tholen's taxonomic class \citep{Tholen89} and diameter (Jet Propulsion
Laboratory database, JPL\footnote{http://www.jpl.nasa.gov/}; Near Earth Objects Dynamic Site,
NeoDys\footnote{http://newton.dm.unipi.it/cgi-bin/neodys/neoibo}) are also listed in Table~\ref{Table2}.

The potential of the data published in the present paper can be fully exploited when
combined with other observations from different apparitions to derive pole positions
and shape models. For this reason the present data are completely available upon
request.

\begin{scriptsize}
\begin{longtable}{lcccccc}

\caption{Synodic rotational period value, period reliability code \citep{Harris83},  lightcurve amplitude, mean $B-V$
         colour index, taxonomic class \citep{Tholen89}, and diameter (JPL and NeoDys database) of the observed asteroids.}\\

\hline
\hline
\noalign{\smallskip}
Asteroid          &     $P_{Syn}$    &   Reliability    &       Light. Amp.    &     $B-V$     & Taxonomic & Diameter \\

                  &      (hours)     &Code\,$^{\dagger}$&         (mag)        &     (mag)     &   Class   &  (km)    \\
\noalign{\smallskip}
\hline
\endfirsthead

\caption{Continued.}\\
\hline
\hline
\noalign{\smallskip}
Asteroid          &     $P_{Syn}$    &   Reliability    &      Light. Amp      &     $B-V$     & Taxonomic & Diameter \\

                  &      (hours)     &Code\,$^{\dagger}$&         (mag)        &     (mag)     &   Class   &  (km)    \\
\noalign{\smallskip}
\hline
\noalign{\smallskip}
\noalign{\smallskip}
\endhead

\hline
\endfoot

\hline
\endlastfoot

  27 $Euterpe$    & 10.377$\pm$0.008 &         2        & 0.14$\pm$0.02\,$^a$  & 0.84$\pm$0.02 &     S     &   96  \\

 173 $Ino$\,$^\ddag$  &  6.113$\pm$0.002 &         3        & 0.13$\pm$0.01\,$^a$  & 0.71$\pm$0.01 &     C     &  154  \\

 173 $Ino$\,$^{\ddag\ddag}$  &  6.111$\pm$0.002 &         3        & 0.14$\pm$0.01\,$^a$  & 0.71$\pm$0.01 &           &       \\

 182 $Elsa$       &  80.23$\pm$0.08  &         3        & 0.69$\pm$0.02\,$^a$  & 0.86$\pm$0.01 &     S     &   44  \\

 539 $Pamina$     &      --          &         1        &        --            & 0.71$\pm$0.01 &     --    &   54  \\

 849 $Ara$        & 4.117$\pm$0.001  &         3        & 0.26$\pm$0.01\,$^a$  & 0.70$\pm$0.01 &     M     &   62  \\

 984 $Gretia$     & 5.780$\pm$0.001  &         3        & 0.66$\pm$0.02\,$^b$  &      --       &     --    &   32  \\

1727 $Mette$\,$^*$& 2.981$\pm$0.001  &         3        & 0.26$\pm$0.01\,$^b$  &      --       &     S     &    7  \\

1727 $Mette$\,$^{**}$& 2.981$\pm$0.002  &         3        & 0.22$\pm$0.01\,$^a$  &      --       &           &       \\

1727 $Mette$\,$^{**}$& 2.981$\pm$0.002  &         3        & 0.20$\pm$0.01\,$^c$  &      --       &           &       \\

3199 $Nefertiti$  & 3.021$\pm$0.002  &         3        & 0.19$\pm$0.02\,$^b$  &      --       &     S     &    2       \\

2004\,UE            &   5.6$\pm$0.2    &         2        & 0.98$\pm$0.02\,$^c$  &      --       &     --    &  0.17-0.38 \\

\label{Table2}
\end{longtable}
\vspace{-1.0 cm}
$^{\dagger}$ Meaning of the reliability code: 1) Result based on fragmentary lightcurves. Insufficient data to
estimate a period value; 2) the result is based on less than full coverage, so that the period may be wrong
by $30\%$ or so; 3) a sure result with no ambiguity and full lightcurve coverage.\\
$^{\ddag}$ December 1996; $^{\ddag\ddag}$ July 2004; $^{*}$ January-February 2003; $^{**}$ April 2003.\\
$^{a}$ $V$ magnitude; $^{b}$ unfiltered magnitude; $^{c}$ $R$ magnitude.
\end{scriptsize}

\subsection{27 Euterpe}
\label{Sec:Euterpe}

The first determination of the synodic rotational period of 27 {\it Euterpe}
was reported by \citet{Chang62}, who observed this asteroid during
three nights in December 1961 and January 1962, obtaining a value of
$P_{syn}=8.500~h$. Their complete lightcurve shows two asymmetrical maxima
and minima having different levels, with an amplitude of $\sim0.15$~mag.
Further photometric observations of 27~{\it Euterpe} were performed by
\citet{Lagerkvist88}, who derived the slope parameter $G$ and the absolute
magnitude $H$, and by \citet{Denchev98}, who observed this asteroid only
for four hours on January 22, 1991 through the $U$~filter, reporting an amplitude
of $0.12$~mag. During the 2001 apparition, Stephens et al. (2001) estimated a longer
synodic rotational period of $10.410\pm0.002~h$, obtaining an ``\emph{interesting
lightcurve with a single strong peak confirmed by multiple nights of observations
and a very weak secondary peak}''.

\begin{figure}
   \centering
   \resizebox{12cm}{!}{\includegraphics[draft=false]{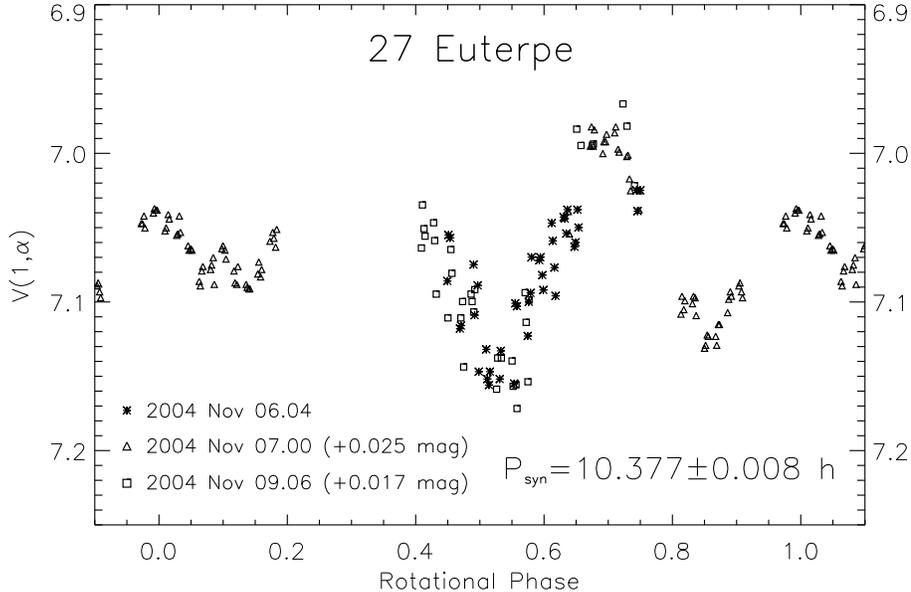}}
   \caption{Composite $V(1,\alpha)$ lightcurve of 27~{\it Euterpe} vs. rotational phase. The zero phase corresponds
   to JD~$2453317.0$ (corrected for light-travel time). Different symbols refer to different observing nights.
   The magnitude shifts derived by applying the Fourier analysis method to each night lightcurve in order to obtain
   the composite curve are reported in brackets after the dates. The rotational phases were computed according to the
   synodic rotational period reported in the lower right corner of the Figure and in Table~\ref{Table2}.}
   \label{Fig:Euterpe-LC}
   \vspace{0.6cm}
\end{figure}

Our first observations of 27~{\it Euterpe}, performed in November 1993 at
CT station, are too few to give a reliable estimation of the rotational
period value. The derived mean $B-V$ colour index is $0.83\pm0.02$~mag. We
Again, we observed this asteroid from the same site during three nights in November
2004, obtaining a lightcurve covering about $80$\,\% of the rotational phase
(Figure~\ref{Fig:Euterpe-LC}). The derived value of the synodic rotational period is
$10.377\pm0.008~h$, slightly shorter than the Stephens et al.'s (2001) one.
The amplitude of the lightcurve is $0.14\pm0.02$~mag; the mean
$B-V$ colour index is $0.85\pm0.02$~mag, in agreement with our previous determination.
If a maximum were located between the 0.2 - 0.4 phase interval, as
it seems to be present both in ours and Stephens et al.'s (2001) lightcurve,
27~{\it Euterpe} would have a three maxima and minima lightcurve.

\subsection{173 Ino}
\label{Sec:Ino}

The first observations of this C-type asteroid 
are by \citet{Lagerkvist78} and \citet{Schober78}. The former detected photometric
variations but with no indication of both rotational period and lightcurve
amplitude; the latter observed 173~{\it Ino} during three nights in 1977, obtaining
an asymmetric lightcurve with the extrema at different level and an amplitude of
$0.04$~mag. The best fit of the photometric data was obtained using a synodic
rotational period of $5.93\pm0.01~h$. Using Schober's period, 
\citet{DiMartinoCacciatori84} covered about $70$\,\% of the rotational phase, deriving an
amplitude of about $0.11$~mag. A longer period was found by \citet{Debehogne90}
and by \citet{Erikson90}, who reported $P_{syn}=6.15\pm0.02~h$ and $P_{syn}=6.11\pm0.06~h$,
respectively.
\citet{Michalowski93} and \citet{DeAngelis95}, using the lightcurves from the 1977, 1983,
and 1988 oppositions, computed the spin axis direction, triaxial shape, and sense of
rotation of 173~{\it Ino}. The former author also derived the sidereal rotational period
$P_{sid}=0.256798~d$ and a new value of the diameter ($D = 159$~km) of this asteroid.
\citet{Michalowski05} performed additional photometric observation during the 1998,
1999, and 2002 apparitions, improving both the shape model and pole orientations, and
deriving a new value of the sidereal rotational period ($P_{sid}=0.2548546\pm0.0000004~d$).

\begin{figure}
   \centering
   \resizebox{12cm}{!}{\includegraphics[draft=false]{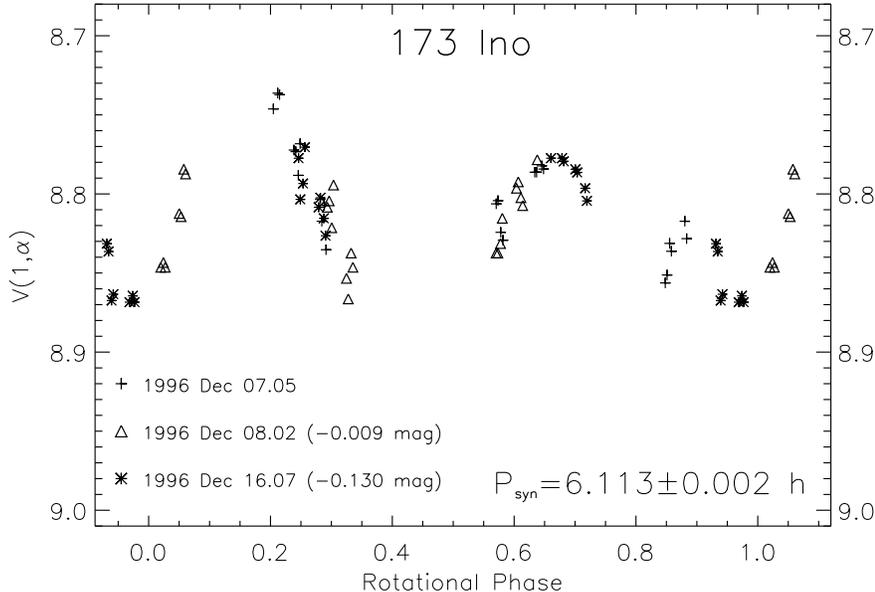}}
   \caption{Composite $V(1,\alpha)$ lightcurve of 173 {\it Ino} vs. rotational phase as derived from the
            photometric data collected in December 1996. The zero phase corresponds to JD~$2450430.0$
            (corrected for light-travel time). See the caption of Figure~\ref{Fig:Euterpe-LC} for further details.}
   \label{Fig:Ino-1996}
   \vspace{0.6cm}
\end{figure}

We observed 173~{\it Ino} during three nights in December 1996 at CT station
(Figure~\ref{Fig:Ino-1996}), deriving a $B-V$ colour index of $0.71\pm0.01$~mag,
in good agreement with previous determinations \citep{Tedesco89}. 
According to a synodic rotational period of $6.113\pm0.002~h$, we obtained a two
maxima and two minima symmetric lightcurve with an amplitude of $\sim0.13\pm0.01$~mag.

\begin{figure}
   \centering
   \resizebox{12cm}{!}{\includegraphics[draft=false]{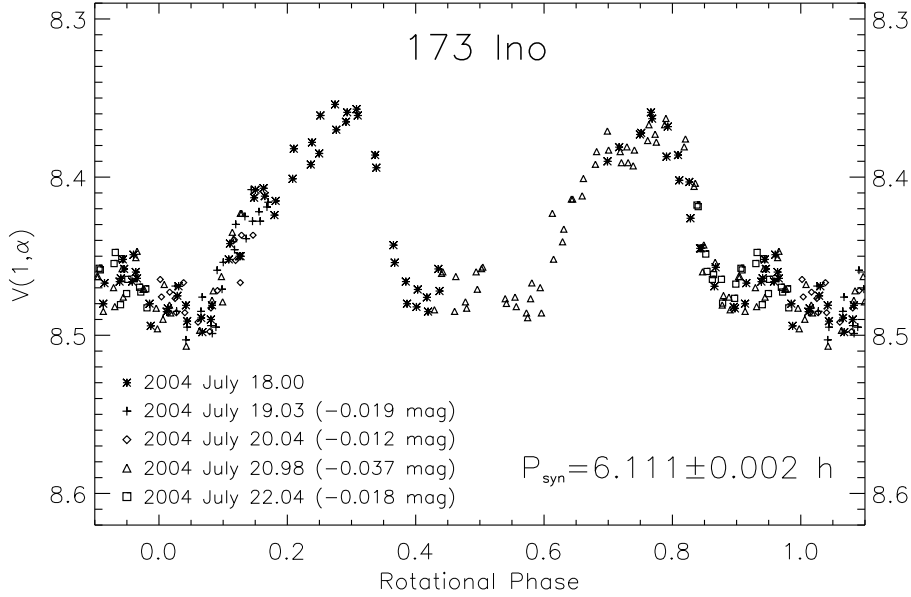}}
   \caption{Composite $V(1,\alpha)$ lightcurve of 173 {\it Ino} vs. rotational phase as derived from the
            photometric data collected in July 2004. The zero phase corresponds to JD~$2453206.0$
            (corrected for light-travel time). See the caption of Figure~\ref{Fig:Euterpe-LC} for further details.}
   \label{Fig:Ino-LC2}
   \vspace{0.6cm}
\end{figure}

Further $B$ and $V$ observations were carried out at the same site during five nights in
July 2004. We derived a synodic rotational period of $6.111\pm0.002~h$ and a $B-V$ colour index 
of $0.71\pm0.01$~mag, both in good agreement with our previous determination.
The composite $V(1,\alpha)$ lightcurve shows four symmetric extrema at the same respective
level and an amplitude of $0.14\pm0.01$~mag (Figure~\ref{Fig:Ino-LC2}). The 1996 and 2004
lightcurves show a similar shape and behaviour. Two small jumps seem to appear
during the minima phases in the 2004 lightcurve (Figure~\ref{Fig:Ino-LC2}); one of them
is clearly visible also in the 1996 lightcurve (Figure~\ref{Fig:Ino-1996}).

\subsection{182 Elsa}
\label{Sec:Elsa}

From previous photoelectric observations of 182~{\it Elsa} carried out by \citet{Harris80,Harris92} 
during the 1978 and 1981 apparitions, only an indication of the rotational period value was 
given ($\sim80~h$).

\begin{figure}
   \centering
   \resizebox{12cm}{!}{\includegraphics[draft=false]{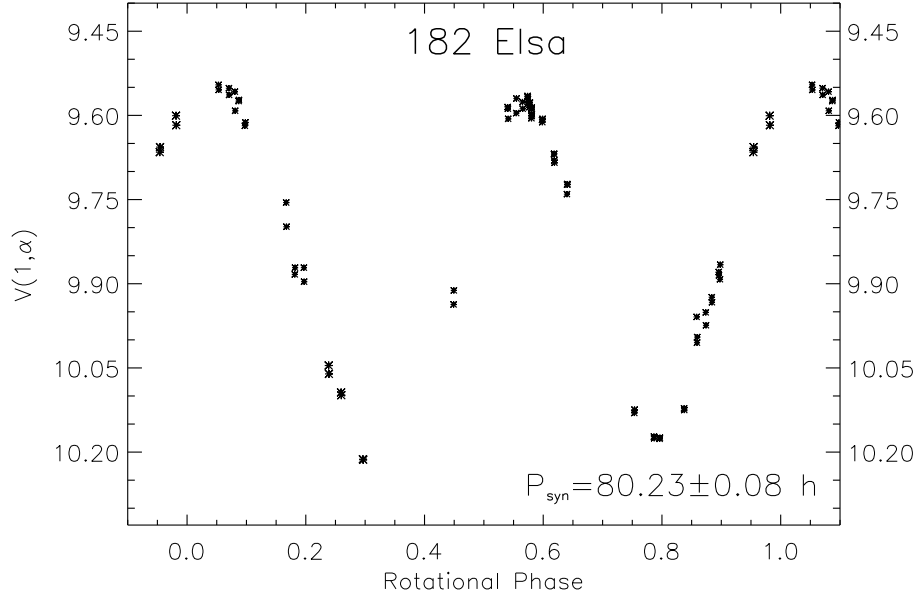}}
   \caption{Composite $V(1,\alpha)$ lightcurve of 182 {\it Elsa} vs. rotational phase as
             derived according to the synodic rotational period reported in the Figure and in Table~\ref{Table2}.
             The zero phase corresponds to JD~$2453271.0$ (corrected for light-travel time).
             Due to the great number of observing nights, the same point-symbol was used for all the nights.
             For the same reason, neither the observing dates nor the nightly magnitude
             shifts were reported in the Figure.}
   \label{Fig:Elsa-LC}
   \vspace{0.6cm}
\end{figure}

Our observations of this S-type asteroid were performed at CT station during eleven nights,
between September and October 2004. We derived a value of the synodic rotational
period of $80.23\pm0.08~h$. The composite $V$ lightcurve shows a nearly symmetrical sinusoidal
trend with an amplitude of $0.69\pm0.02$~mag (Figure~\ref{Fig:Elsa-LC}). The two maxima
have the same height, while the minima show different levels. The mean value of $B-V$ colour
index is $0.86\pm0.01$~mag.

\begin{figure}
   \centering
   \resizebox{12cm}{!}{\includegraphics[draft=false]{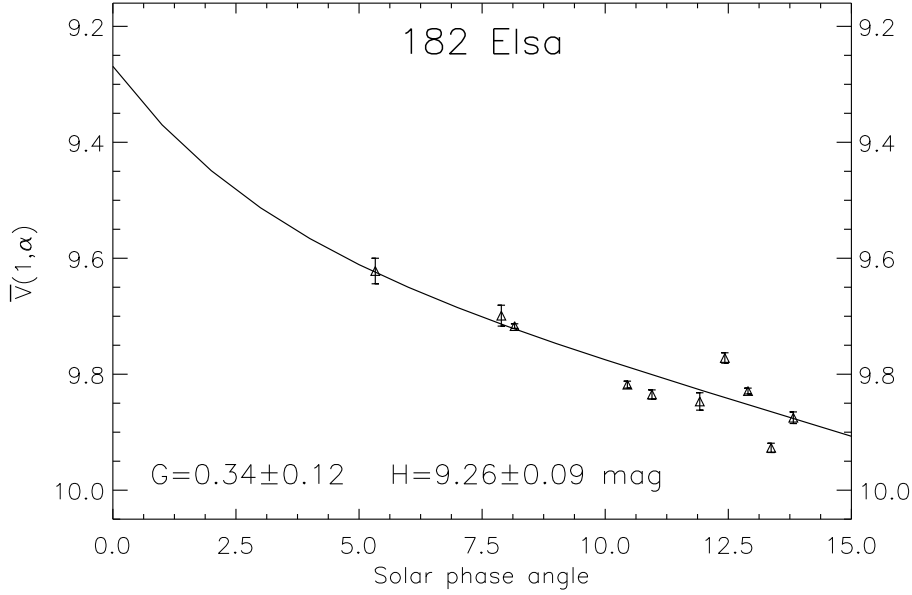}}
   \caption{$H-G$ magnitude relation for 182 {\it Elsa}.}
   \label{Fig:Elsa-HG}
   \vspace{0.6cm}
\end{figure}

During our eleven-night photometric run, the solar phase angle of 182~{\it Elsa} varied
between $\sim5^\circ$ and $14^\circ$. We used the nightly mean reduced magnitudes
$\bar{V}(1,\alpha)$ to obtain a least-squares fit of the $H-G$ magnitude relation, as
described by \citet{Bowell89}. The mean reduced magnitude $\bar{V}(1,\alpha)$ and
fitted phase curve, both vs. phase angle, are plotted in Figure~\ref{Fig:Elsa-HG}. The
best-fitting value for $H$ and $G$ is $9.26\pm0.09$~mag and $0.34\pm0.12$, respectively.
The latter result is consistent with the S-type taxonomic classification of this asteroid.

\subsection{539 Pamina}
\label{Sec:Pamina}

This asteroid was observed only during one night in August 1981 by \citet{Harris92}, 
who obtained a mean $B-V$ colour index of $0.70\pm0.01$~mag, detecting a
magnitude variations but without any indication of a possible rotational period
or amplitude. The first complete lightcurve was obtained by \citet{Pray05} who
derived a synodic rotational period of $13.903\pm0.001~h$, with an amplitude of
$0.10\pm0.01$~mag.

Our observations were carried out during two nights in September 2004 at CT station.
We found a mean $B-V$ colour index value of $0.71\pm0.01$ mag, in good agreement
with the one reported by \citet{Harris92}. Due to the poor sampling of our
data, we were not able to construct a significant composite lightcurve,
even using the period value reported by \citet{Pray05}.

\subsection{849 Ara}
\label{Sec:Ara}

The only published lightcurve of this M-type asteroid was obtained during six observing
nights in May-June 1981 by \citet{Harris92}, who found a synodic rotational period
of $4.11643\pm0.00005~h$.

\begin{figure}
   \centering
   \resizebox{12cm}{!}{\includegraphics[draft=false]{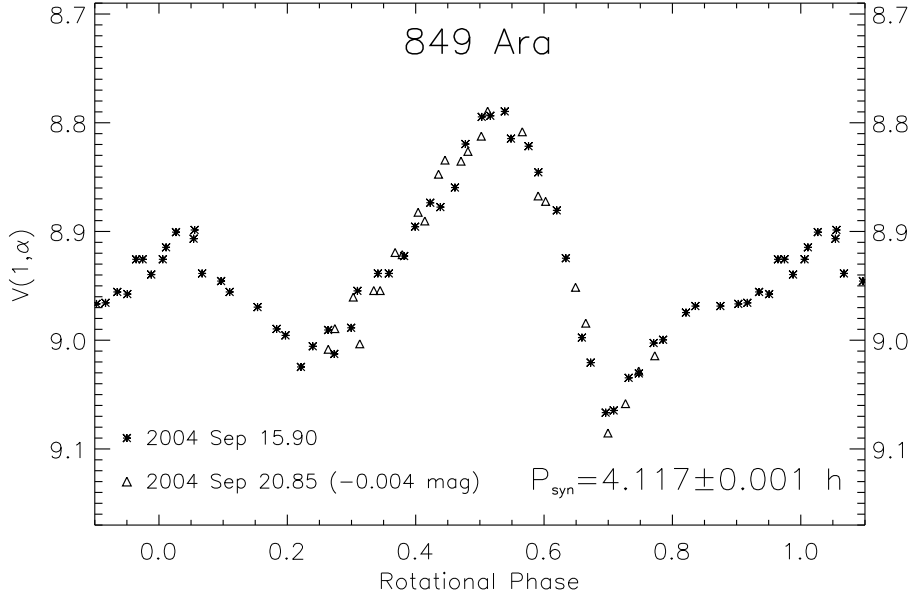}}
   \caption{Composite $V(1,\alpha)$ lightcurve of 849~{\it Ara} vs. rotational phase. The zero
            phase corresponds to JD~$2453266.0$ (corrected for light-travel time). See the caption of
            Figure~\ref{Fig:Euterpe-LC} for further details.}
   \label{Fig:Ara-LC}
   \vspace{0.6cm}
\end{figure}

849~{\it Ara} was observed at CT station on September 15 and 20, 2004.
We derived a synodic rotational period value of $4.117\pm0.001~h$, in very good agreement
with the one reported by \citet{Harris92}. The well-covered composite lightcurve
(Figure~\ref{Fig:Ara-LC}) shows four extrema as well as an amplitude of $0.26\pm0.01$~mag.
The lightcurve behaviour appears irregular, with a linear trend before and after the
highest maximum peak and falling-in and jumps in the remaining lightcurve parts. The mean
$B-V$ colour index is $0.70\pm0.01$~mag, in good agreement with the one reported by
\citet{Tedesco89}.

\subsection{984 Gretia}
\label{Sec:Gretia}

The first lightcurve of 984~{\it Gretia} was obtained from photographic plates by
\citet{vanHouten62} who derived a synodic rotational period of $5.76~h$. This
value was improved to $5.781~h$ by \citet{DiMartino84} and \citet{Piironen94}. 
Additional observations were performed by two of the present authors
\citep{Riccioli01} who obtained a synodic period of $5.560\pm0.018~h$,
i.e. about two hundredth shorter than the one previously published by 
\citet{DiMartino84} and \citet{Piironen94}.

\begin{figure}
   \centering
   \resizebox{12cm}{!}{\includegraphics[draft=false]{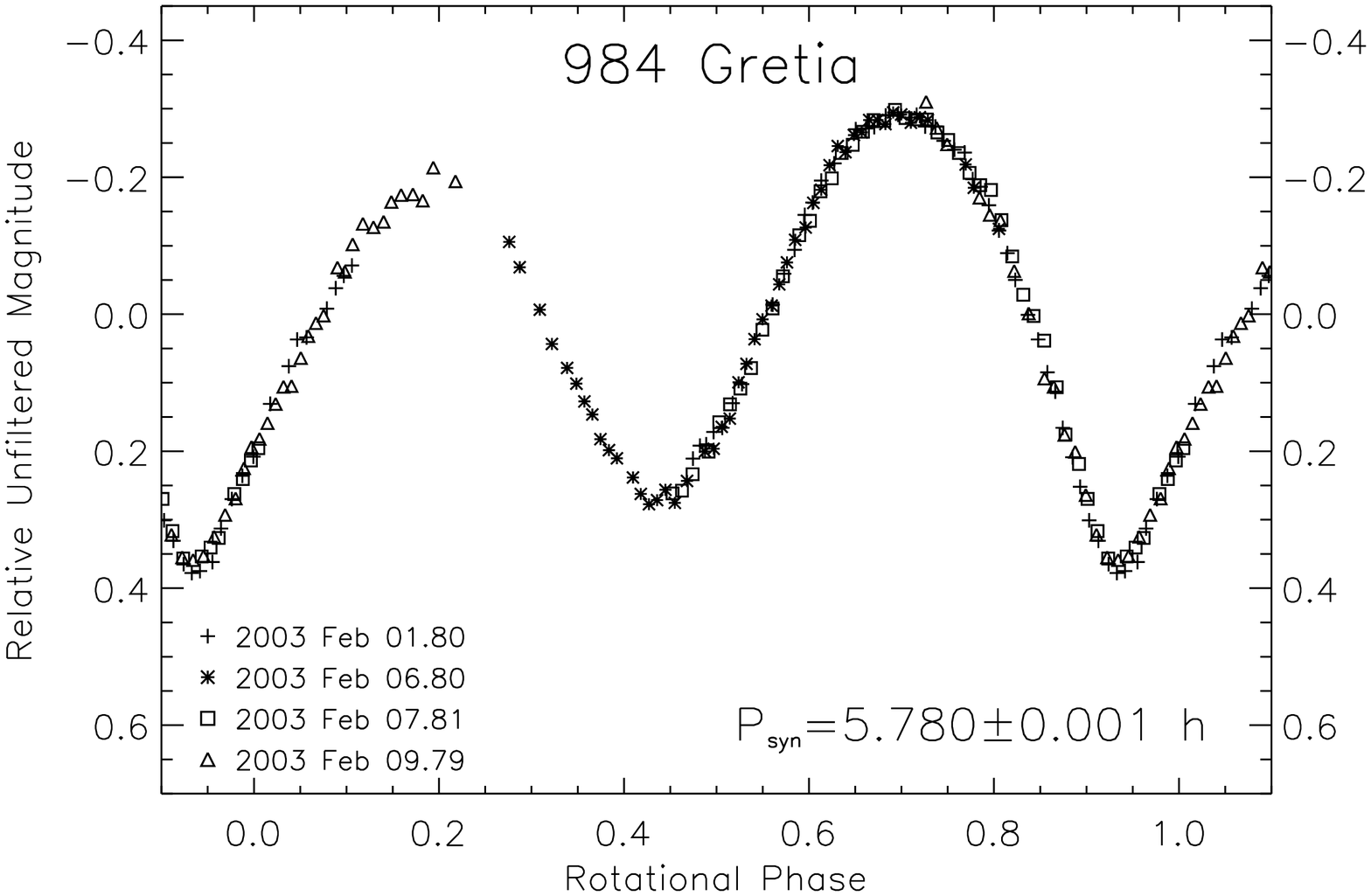}}
   \caption{Unfiltered composite lightcurve of 984 {\it Gretia} vs. rotational phase.
            The zero phase corresponds to JD~$2452675.0$ (corrected for light-travel time).
            See the caption of Figure~\ref{Fig:Euterpe-LC} for further details.}
   \label{Fig:Gretia-LC}
   \vspace{0.6cm}
\end{figure}

The present data of 984~{\it Gretia} were obtained during four nights in the first
decade of February 2003 at PD station using the SCAM-1 CCD camera, without filters.
Figure~\ref{Fig:Gretia-LC} reports the relative magnitude lightcurve, phase folded
with the $5.780\pm0.001~h$ synodic rotational period value, in good agreement with
the one reported by \citet{Piironen94} and \citet{DiMartino84}. The lightcurve
behaviour is regular with four symmetric extrema at different level and an
amplitude of $0.66\pm0.02$~mag.

\subsection{1727 Mette}
\label{Sec:Mette}

The asteroid 1727~{\it Mette} is a relatively small object 
\citep[$D\approx7$~km;][]{WisniewskiMcMillan87}, belonging to the Hungaria zone, 
as defined by \citet{Zellner85}. Previous CCD photometric observations were 
reported by \citet{WisniewskiMcMillan87}, \citet{Prokofeva92}, and \citet{Sarneczky99},
who derived unreliable synodic rotational periods of $2.637\pm0.004~h$ and
$3.22\pm0.22~h$, on the basis of poor-sampled and incomplete lightcurves.

\begin{figure}
   \centering
   \resizebox{12cm}{!}{\includegraphics[draft=false]{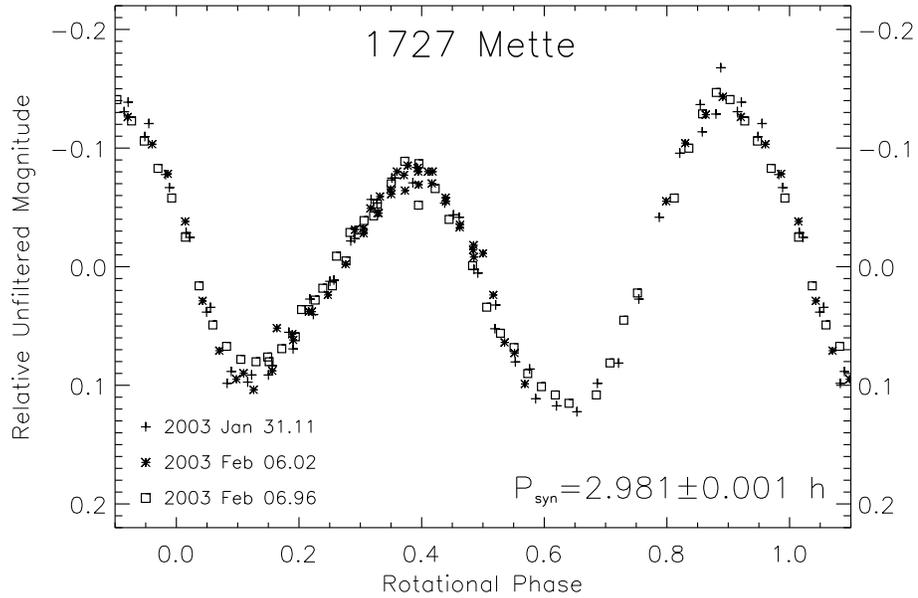}}
   \caption{Unfiltered composite lightcurve of 1727 {\it Mette} vs. rotational phase. The zero
            phase corresponds to JD~$2452675.0$ (corrected for light-travel time). See the caption of
            Figure~\ref{Fig:Euterpe-LC} for further details.}
   \label{Fig:Mette-LC-SCAM1}
   \vspace{0.6cm}
\end{figure}

We observed 1727~{\it Mette} using the CCD SCAM-1 camera at PD station.
The observations were performed in white-light during three nights in January
and February 2003. The composite lightcurve (Figure~\ref{Fig:Mette-LC-SCAM1})
shows an amplitude of $0.26\pm0.01$~mag and an almost symmetric trend with four
well-defined extrema at different depth and height. We derived a synodic
rotational period of $2.981\pm0.001~h$, improving the previous determinations.

In order to confirm our result, further $V$ and $R$ photometric observations
were carried out during five nights in April 2003, at the same observing site using the
CCD ITANET camera. We derived a synodic rotational period of $2.981\pm0.002~h$,
in perfect agreement with our previous value.

\begin{figure}
   \centering
   \resizebox{12cm}{!}{\includegraphics[draft=false]{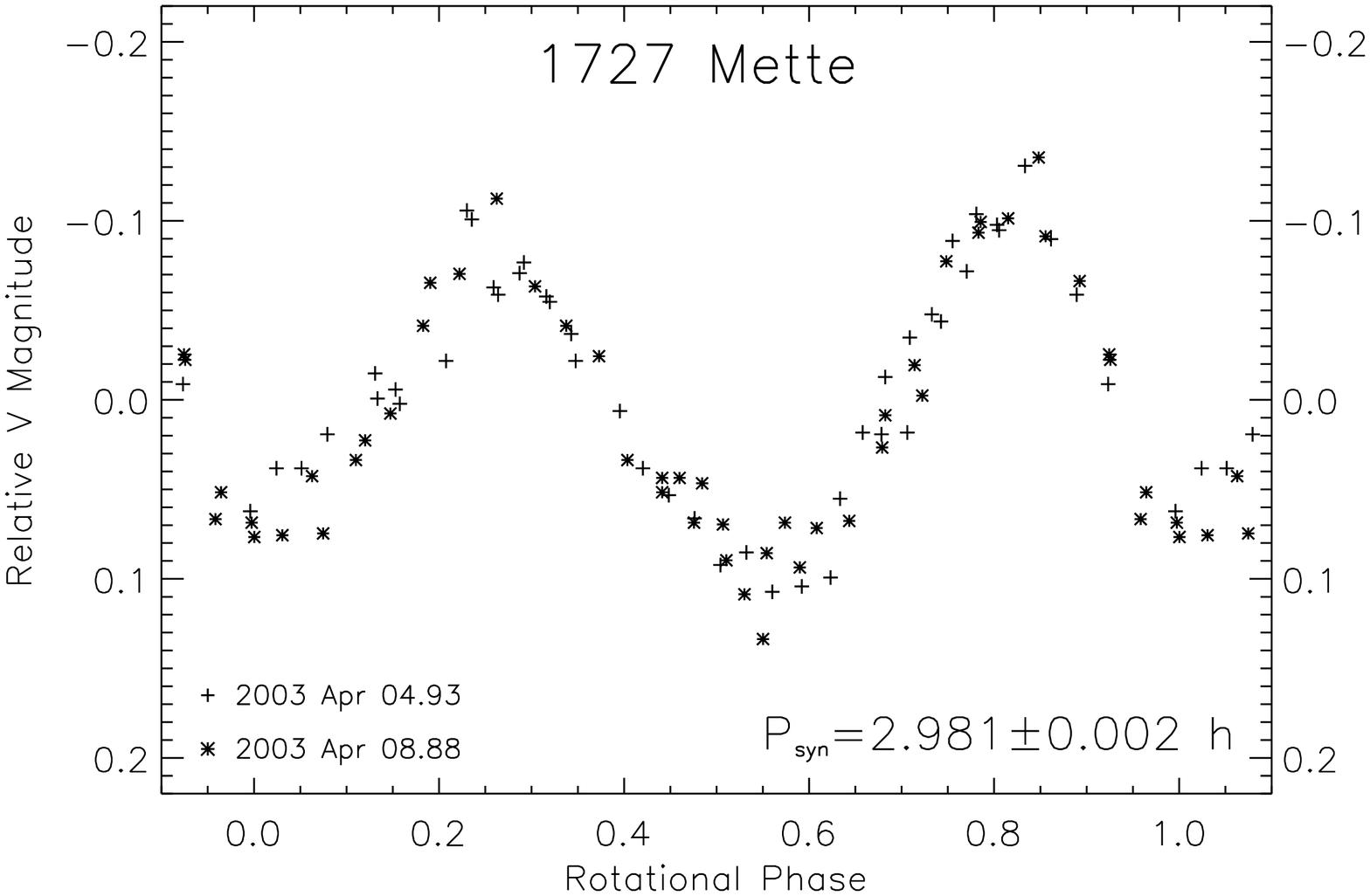}}
   \vspace{0.5cm}
   \centering
   \resizebox{12cm}{!}{\includegraphics[draft=false]{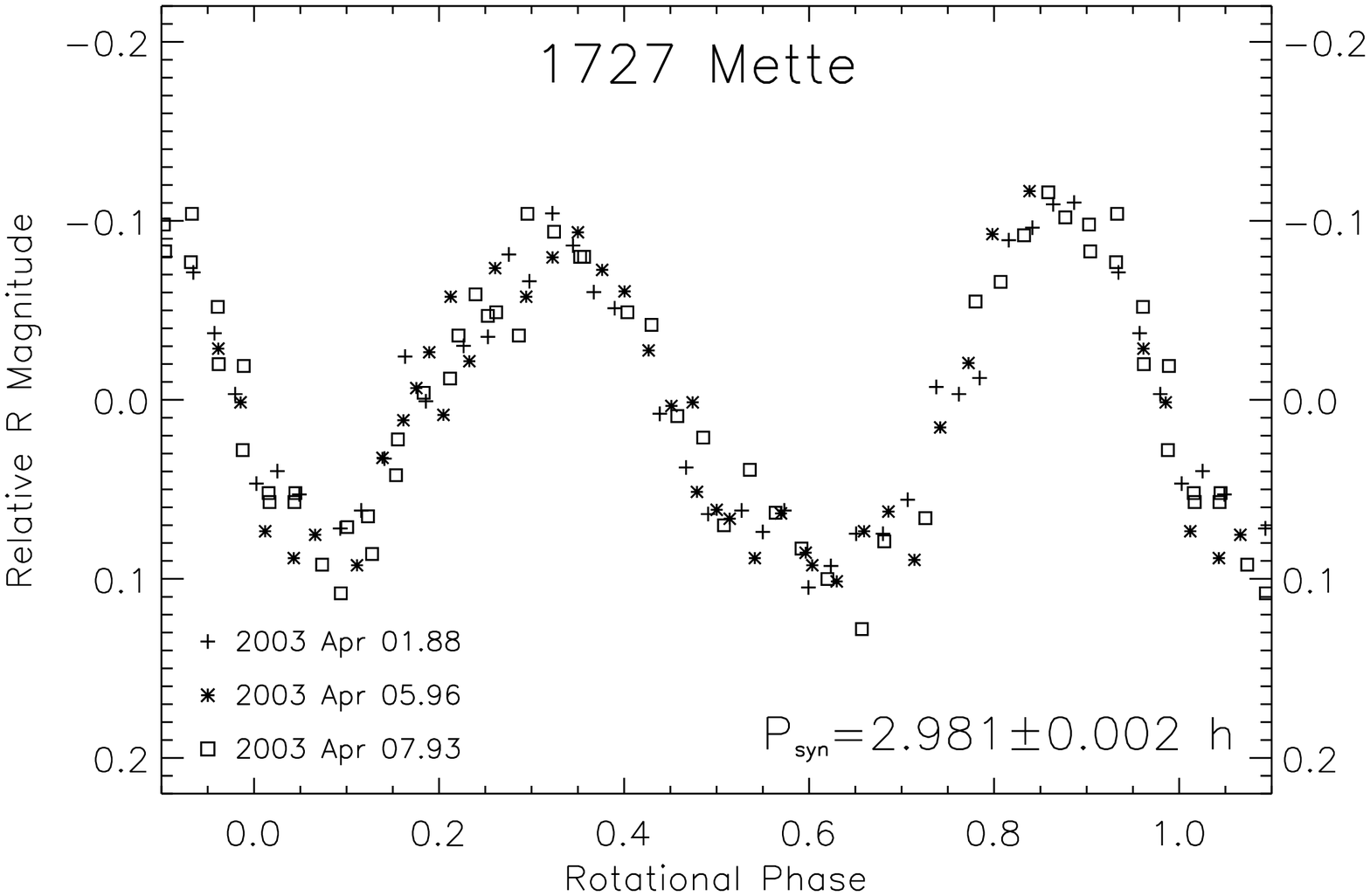}}
   \caption{Composite $\Delta V$ and $\Delta R$ lightcurve of 1727~{\it Mette} vs. rotational phase.
            The zero phase corresponds to JD~$2452736.0$ and JD~$2452735.0$, respectively (corrected
            for light-travel time). See the caption of Figure~\ref{Fig:Euterpe-LC} for further details.}
   \label{Fig:Mette-LC-ITANET}
   \vspace{0.6cm}
\end{figure}

The $V$ and $R$ lightcurve are plotted in Figure~\ref{Fig:Mette-LC-ITANET}.
They both show a trend similar to that of Figure~\ref{Fig:Mette-LC-SCAM1} as well as
four well-defined extrema at slightly different levels. The $V$ and $R$ lightcurve amplitudes
are $0.22\pm0.01$ and $0.20\pm0.01$~mag, respectively. Despite the lightcurve amplitude usually
increases with increasing phase angle, the latter values are slightly lower than our
previous determination. This might be due to a variation of the aspect angle between the two
observing runs. In fact the viewing/illumination geometry can also play a role in changing the
amplitude of the lightcurve.

\subsection{3199 Nefertiti}
\label{Sec:Nefertiti}

The first photometric observations of this S-Amor NEA were performed by \citet{Harris85} 
who obtained a synodic rotational period of $3.01~h$ and an amplitude of $0.12$~mag.
During the 1986 apparition \citet{Wisniewski87} derived a shorter value of $2.816~h$.
\citet{Pravec97} using photometric data from the 1982, 1984, 1994, and 1995 apparitions,
obtained a synodic rotational period of $3.0207\pm0.0002~h$, close to the value reported
by \citet{Harris85}. \citet{Kaasalainen04}, thanks to a long-term photometric
data-set spanning from 1982 to 2003, obtained a rotational period value of $3.020167~h$,
as well as the pole position and shape model.

\begin{figure}
   \centering
   \resizebox{12cm}{!}{\includegraphics[draft=false]{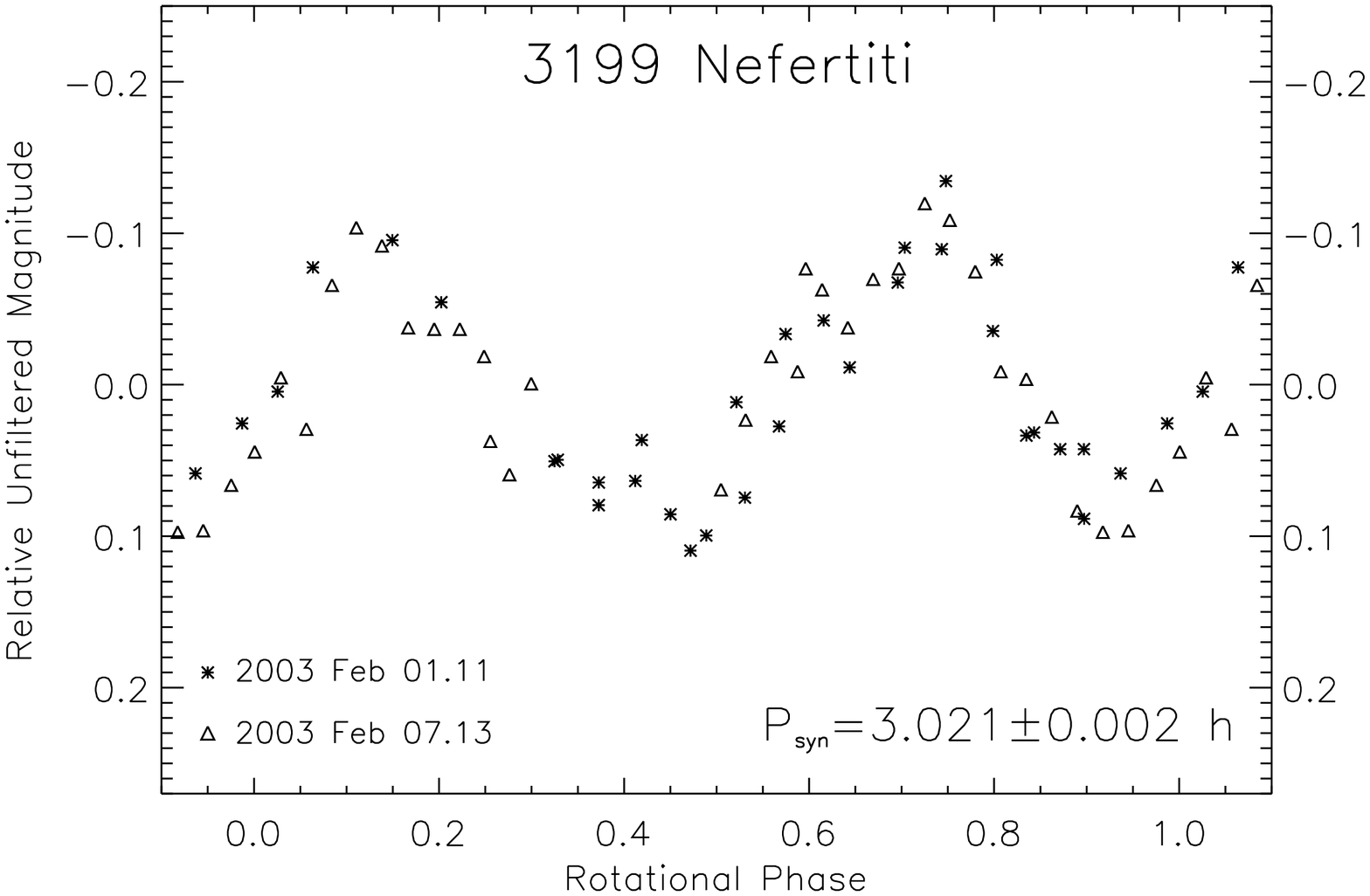}}
   \caption{Composite unfiltered lightcurve of 3199 {\it Nefertiti} vs. rotational phase. The
            zero phase corresponds to JD~$2452675.0$ (corrected for light-travel time). See the
            caption of Figure~\ref{Fig:Euterpe-LC} for further details.}
   \label{Fig:Nefertiti-LC}
   \vspace{0.6cm}
\end{figure}

We observed 3199~{\it Nefertiti} during two nights in January and February 2003 using the
SCAM-1 CCD camera at PD station. The composite unfiltered lightcurve, (Figure~\ref{Fig:Nefertiti-LC}),
shows a nearly sinusoidal trend with four well-defined extrema, as well as an amplitude of
$0.19\pm0.02$~mag. We derived a synodic rotational period of $3.021\pm0.002~h$, in good agreement 
with the values obtained by \citet{Pravec97} and \citet{Kaasalainen04}.

\subsection{2004 UE}
\label{Sec:2004UE}

This small Apollo-type asteroid ($D\sim$ 0.17-0.38~km) 
was discovered by the LINEAR survey in October 2004. We observed this asteroid at PD station
using the CCD ITANET camera. $R$ filter photometric observations were collected during its
close path to the Earth in November, 2004, when the asteroid reached its maximum apparent $V$
brightness of about $14.2$~mag. Unfortunately, we were able to observe 2004~UE only during
one night both due to bad weather conditions and the short time-span of the favorable apparition.

\begin{figure}
\centering
   \resizebox{12cm}{!}{\includegraphics[draft=false]{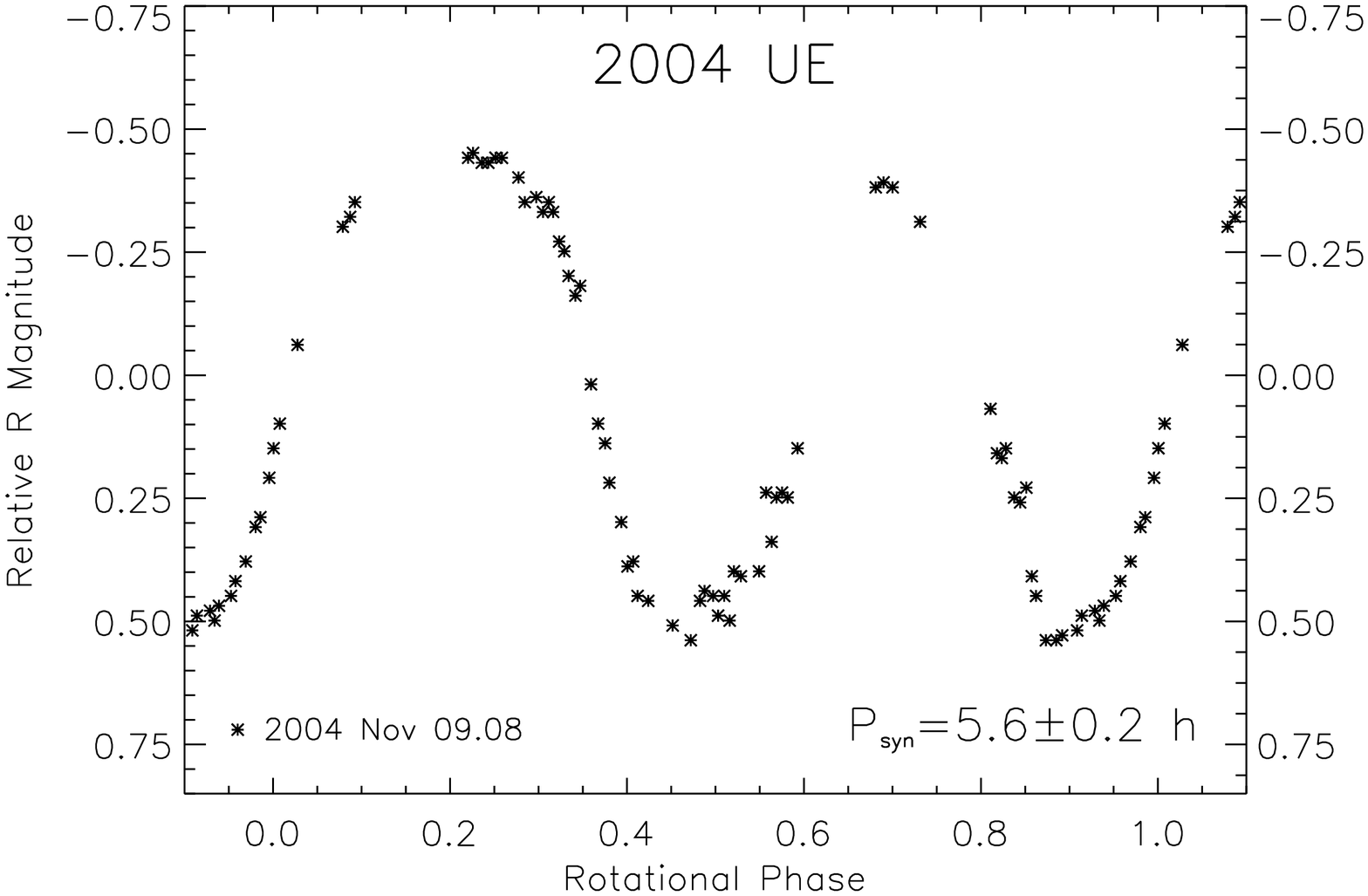}}
   \caption{Composite $\Delta R$ lightcurve of 2004~UE vs. rotational phase. The zero phase
            corresponds to JD~$2453319.0$ (corrected for light-travel time). See the caption of
            Figure~\ref{Fig:Euterpe-LC} for further details.}
   \label{Fig:2004UE-LC}
   \vspace{0.6cm}
\end{figure}

The $R$ lightcurve is presented in Figure~\ref{Fig:2004UE-LC}. We derived a synodic rotational
period value of $5.6\pm0.2~h$ and a magnitude variation of $0.98\pm0.02$~mag. The lightcurve
shows a smoothed trend with four nearly symmetrical extrema. The large amplitude value and the
regular trend of the lightcurve might indicate an elongated shape without large surface
dishomogeneities.

\newpage

{\bf Acknowledgements}

We thank the anonymous referee for his/her careful reading, useful
comments, and suggestions that helped us to improve the manuscript.
We thank Padova Astronomical Observatory for the use of the
67/92-cm Schmidt telescope. We are kindly grateful to R. Claudi
and to all technical staff for the precious assistance during the
observations. The authors also wish to thank Luigia Santagati for
the english revision of the text. Davide Gandolfi is grateful to
Menadora Barcellona and Salvatore Spezzi whom memory supported him
during the preparation of this paper.





\begin{thebibliography}{}


\bibitem[Bertin \& Arnouts(1996)]{Bertin96} Bertin E. \& Arnouts S. 1996,  A\&AS 117, 393
\bibitem[Blanco et al.(2004)]{Blanco04} Blanco C., Bonanno G., Belluso M., Bruno P., Cal\`{\i} A., Gandolfi D. and Timpanaro M.C. 2004, Proceedings of the {\it V Convegno di Scienze Planetarie}, Gallipoli (Lecce), September 15-19, 2003, 131
\bibitem[Blanco et al.(1968)]{Blanco68} Blanco V.M., Demers S., Douglass G.G. and Fitzgerald M.P. 1968, PUSNO~21
\bibitem[Blanco \& Catalano(1979)]{Blanco79} Blanco C. and Catalano S. 1979, Icarus 40, 359
\bibitem[Blanco \& Riccioli(1998)]{Blanco98} Blanco C. \& Riccioli D. 1998, A\&AS 131, 385
\bibitem[Bowell et al.(1989)]{Bowell89} Bowell E., Hapke B., Domingue D. et al. 1989, {\it Asteroid II}, University of Arizona Press, Tucson, 524
\bibitem[Chang \& Chang(1962)]{Chang62} Chang Y.C. \& Chang C.S. 1962, Acta Astron. Sin. 10, 101
\bibitem[De Angelis(1995)]{DeAngelis95} De Angelis G. 1995, P\&SS 43, 649
\bibitem[Debehogne et al.(1990)]{Debehogne90} Debehogne H., Lagerkvist C.-I., Magnusson P. and Hahn G. 1990, Proceedings of {\it Asteroids, Comets, Meteors III}, Astronomical Observatory of the Uppsala University, June 12-16, 1989, Uppsala, 45
\bibitem[Denchev et al.(1998)]{Denchev98} Denchev P., Magnusson P. and Donchev Z. 1998, P\&SS 46, 673
\bibitem[Di Martino \& Cacciatori(1984)]{DiMartinoCacciatori84} Di Martino M. \& Cacciatori S. 1984, Icarus 60, 75
\bibitem[Di Martino(1984)]{DiMartino84} Di Martino M. 1984, Icarus 60, 541
\bibitem[Di Martino et al.(1994)]{DiMartino94} Di Martino M., Blanco C., Riccioli D. and De Sanctis G. 1994, Icarus 107, 269
\bibitem[Erikson(1990)]{Erikson90} Erikson A. 1990, Proceedings of {\it Asteroids, Comets, Meteors III}, Astronomical Observatory of the Uppsala University, June 12-16, 1989, Uppsala, 55
\bibitem[Gandolfi et al.(2006)]{Gandolfi06} Gandolfi D., Blanco C., Bonanno G., et al 2006,  MSAIS 9, 180
\bibitem[Harris et al.(1980)]{Harris80} Harris A.W., Young J.W., Scaltriti F. and Zappal\`a V. 1980, Icarus 41, 316
\bibitem[Harris \& Young(1983)]{Harris83} Harris A.W. \& Young J.W. 1983, Icarus 54, 59
\bibitem[Harris \& Young(1985)]{Harris85} Harris A.W. \& Young J.W. 1985, BAAS 17, 726
\bibitem[Harris et al.(1989)]{Harris89} Harris A.W., Young J.W., Bowell E. et al. 1989, Icarus 77, 171
\bibitem[Harris et al.(1992)]{Harris92} Harris A.W., Young J.W., Dockweiler T. et al. 1992, Icarus 95, 115
\bibitem[Harris(2006)]{Harris06} Harris A.W. 2006, Proceedings of IAU Symposium 229: {\it Asteroids, Comets, Meteors}, B\'uzios, Rio de Janeiro, August 7-12, 2005, pp. 449-463
\bibitem[Lagerkvist(1978)]{Lagerkvist78} Lagerkvist C.-I. 1978, A\&AS 31, 361
\bibitem[Lagerkvist et al.(1988)]{Lagerkvist88} Lagerkvist C.-I., Magnusson P., Williams I.P. et al. 1988, A\&AS 73, 395
\bibitem[Landolt(1992)]{Landolt92} Landolt A.U. 1992 AJ, 104, 340
\bibitem[Kaasalainen et al(2004)]{Kaasalainen04} Kaasalainen M., Pravec P., Krugly Y.N. et al. 2004, Icarus 167, 178
\bibitem[Mermilliod et al.(1997)]{Mermilliod97} Mermilliod J.-C., Mermilliod M. and Hauck B. 1997, A\&AS 124, 349
\bibitem[Michalowski(1993)]{Michalowski93} Michalowski T. 1993, Icarus 106, 563
\bibitem[Michalowski et al.(2005)]{Michalowski05} Michalowski T., Kaasalainen M., Marciniak A. at al. 2005, A\&A 443, 329
\bibitem[Piironen et al.(1994)]{Piironen94} Piironen J., Bowell E., Erikson A. and Magnusson P. 1994, A\&AS 106, 587
\bibitem[Pravec et al.(1997)]{Pravec97} Pravec P., Wolf M., Sarounova L. et al. 1997, Icarus 130, 275
\bibitem[Pray(2005)]{Pray05} Pray P. 2005, MPBu 32, 8
\bibitem[Prokof'eva et al.(1992)]{Prokofeva92} Prokof'eva V.V., Demchik M.I. and Golub A.I. 1992, AVest 26, 68
\bibitem[Riccioli et al.(2001)]{Riccioli01} Riccioli D., Blanco C. and Cigna M. 2001, P\&SS 49, 657
\bibitem[Sarneczky et al.(1999)]{Sarneczky99} Sarneczky K., Szabo G. and Kiss L.L. 1999, A\&AS 137, 363
\bibitem[Schober(1978)]{Schober78} Schober H.J. 1978, A\&AS 34, 377
\bibitem[Stephens et al.(2001)]{Stephens01} Stephens R.D., Malcolm G., Koff R.A., Brincat S.M. and Warner B. 2001, MPBu 28, 1
\bibitem[Tedesco(1989)]{Tedesco89} Tedesco E.F. 1989, {\it Asteroid II}, University of Arizona Press, Tucson, 1090
\bibitem[Tholen(1989)]{Tholen89} Tholen D.J. 1989, {\it Asteroid II}, University of Arizona Press, Tucson, 1139
\bibitem[van Houten(1962)]{vanHouten62} van Houten C.J. 1962, BAN 16, 160
\bibitem[Wisniewski(1987)]{Wisniewski87} Wisniewski W.Z. 1987, Icarus 70, 566
\bibitem[Wisniewski \& McMillan(1987)]{WisniewskiMcMillan87} Wisniewski W.Z. \& McMillan R.S. 1987, AJ 93, 1264
\bibitem[Zellner et al.(1985)]{Zellner85} Zellner B., Thirungari A. and Bender D. 1985, Icarus 62, 505

\end{thebibliography}
\end{document}